\documentclass[aps,prl,twocolumn,showpacs,byrevtex]{revtex4}

\usepackage{times,mathptm,amsmath,amssymb}
\usepackage[dvips]{graphicx,color}

\begin{document}
\title{Mott- versus Slater-type Insulating Nature of Two-Dimensional Sn Atom Lattice on SiC(0001)}
\author{Seho Yi$^1$, Hunpyo Lee$^2$, Jin-Ho Choi$^3$, and Jun-Hyung Cho$^{1*}$}
\affiliation{$^1$ Department of Physics and Research Institute for Natural Sciences, Hanyang University, 17 Haengdang-Dong, Seongdong-Ku, Seoul 133-791, Korea \\
$^2$ School of General Studies, Kangwon National University, 346 Jungang-ro, Samcheok-si, Kangwon-do, Korea \\
$^3$ Department of Energy and Materials Engineering and Advanced Energy and Electronic Materials Research Center, Dongguk University-Seoul, Seoul 100-715, Korea
}
\date{\today}

\begin{abstract}
Semiconductor surfaces with narrow surface bands provide unique playgrounds to search for Mott-insulating state. Recently, a combined experimental and theoretical study [Phys. Rev. Lett. {\bf 114}, 247602 (2015)] of the two-dimensional (2D) Sn atom lattice on a wide-gap SiC(0001) substrate proposed a Mott-type insulator driven by strong on-site Coulomb repulsion $U$. Our systematic density-functional theory (DFT) study with local, semilocal, and hybrid exchange-correlation functionals shows that the Sn dangling-bond state largely hybridizes with the substrate Si 3$p$ and C 2$p$ states to split into three surface bands due to the crystal field. Such a hybridization gives rise to the stabilization of the antiferromagnetic order via superexchange interactions. The band gap and the density of states predicted by the hybrid DFT calculation agree well with photoemission data. Our findings not only suggest that the Sn/SiC(0001) system can be represented as a Slater-type insulator driven by long-range magnetism, but also have an implication that taking into account long-range interactions beyond the on-site interaction would be of importance for properly describing the insulating nature of Sn/SiC(0001).
\end{abstract}

\maketitle

Search for Mott-insulating state driven by short-range electron correlations has long been one of the most challenging issues in condensed matter physics~\cite {ima,meng}. Since the electrons in two-dimensional (2D) atom lattices can experience strong on-site Coulomb repulsion $U$ due to their reduced screening, metal overlayers on semiconductor substrates have attracted much attention for the realization of a Mott-Hubbard insulator~\cite{nor,ani,cor,mod,pro,li,li2}, where $U$ splits a half-filled band into lower and upper Hubbard bands. For example, the 1/3-monolayer adsorption of Sn atoms on the Si(111) or Ge(111) surface produces the ${\sqrt{3}}{\times}{\sqrt{3}}$ reconstruction in which all the dangling bonds (DBs) of underlying Si or Ge surface atoms are saturated to leave a single DB on each Sn atom~\cite{cor,mod,pro,li,li2,mor,car,bal,lee1,lee2}. Such Sn-overlayer systems with a half-filled band have been considered as an ideal playground for investigating 2D correlated physics on the ${\sqrt{3}}{\times}{\sqrt{3}}$ triangular lattice~\cite{cor,mod,pro,li,li2}. However, the nature of the insulating ground state in Sn/Si(111) or Sn/Ge(111) has become a controversial issue whether the gap formation is driven by strong Coulomb interactions (Mott-type insulator)~\cite{cor,mod,pro,li,li2} or by long-range magnetic order (Slater-type insulator)~\cite{lee1,lee2}.

To realize a significantly reduced adatom-substrate hybridization as well as a strongly suppressed screening, Glass $et$ $al$.~\cite{gla} fabricated the ${\sqrt{3}}{\times}{\sqrt{3}}$ phase of Sn overlayer on a wide-gap SiC(0001) substrate (see Fig. 1). In their photoemission experiment on the Sn/SiC(0001) surface system, Glass $et$ $al$. observed a large energy gap of ${\sim}$2 eV. To account for the origin of such an insulating phase, Glass $et$ $al$. performed the combined density-functional theory and dynamical mean-field theory (DFT + DMFT) calculations for a single-band Hubbard model that includes only the on-site Coulomb repulsion, and reproduced the experimentally observed insulating gap. Meanwhile, their spin-polarized DFT calculation~\cite{gla} with the local density approximation (LDA) predicted a small energy gap of ${\sim}$0.1 eV for the collinear antiferromagnetic (AFM) ordering. Based on these results, Glass $et$ $al$. interpreted the Sn/SiC(0001) surface system as a pronounced Mott-type insulator. However, the theoretical analysis of Glass $et$ $al$.~\cite{gla} leading to the Mott-insulating scenario raises the following questions: (i) Is the single-band Hubbard model employed in the previous DFT + DMFT calculation~\cite{gla} suitable for describing the insulating nature of the Sn/SiC(0001) system? and (ii) Does the LDA accurately predict the insulating gap formed by the AFM order?

\begin{figure}[ht]
\centering{ \includegraphics[width=8.0cm]{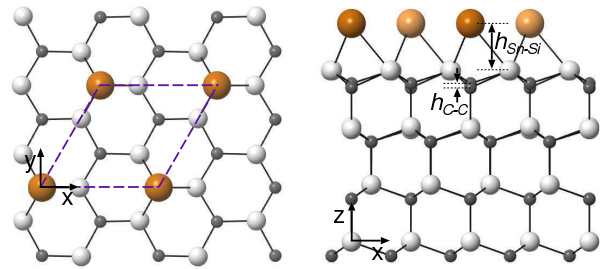} }
\caption{(Color online) Top (left) and side (right) views of the structure of Sn/SiC(0001). The dashed line indicates the ${\sqrt{3}}{\times}{\sqrt{3}}$ unit cell. The {\bf x}, {\bf y}, and {\bf z} axes point along the [1000], [01${\overline{1}}$0], and [0001] directions, respectively. The large, medium, and small circles represent Sn, Si, and C atoms, respectively. For distinction, Sn atoms on the different $y$ positions are drawn with dark and bright circles in the side view.  }
\end{figure}

In this Letter, we investigate the nature of the insulating ground state of Sn/SiC(0001) by using the systematic DFT calculations with the LDA, semilocal (GGA), and hybrid exchange-correlation functionals as well as the LDA + DMFT calculation. All of the DFT calculations predict the AFM ground state, but the calculated band gap largely depends on the employed exchange-correlation functionals. Specifically, the hybrid DFT results for the band gap and the density of states (DOS) agree well with photoemission data. It is revealed that the Sn 5$p_x$, 5$p_y$, and 5$p_z$ orbitals largely hybridize with the substrate Si 3$p_z$ and C 2$p_z$ orbitals, leading to three surface bands due to the crystal-field splitting. Such an unexpectedly large hybridization between the Sn DB state and the substrate states not only facilitates the superexchange interactions between neighboring Sn atoms to stabilize the AFM order, but also implies that long-range interactions beyond the on-site interaction should be taken into account for properly describing the insulating nature of Sn/SiC(0001). The present results suggest that the Sn/SiC(0001) surface system can be more represented as a Slater-type insulator via long-range magnetism rather than the previously~\cite{gla} proposed Mott-type insulator via strong on-site Coulomb repulsion.

We begin to optimize the atomic structure of the NM ${\sqrt{3}}{\times}{\sqrt{3}}$ structure using the LDA, GGA, and hybrid DFT calculations~\cite{method}. The optimized NM structure obtained using LDA is displayed in Fig. 1. We find that the LDA height difference between the Sn atom and its bonding Si atoms is $h_{\rm Sn-Si}$ = 2.09 {\AA} and that between the first C-layer atoms is $h_{\rm C-C}$ = 0.19 {\AA}, in good agreement with those ($h_{\rm Sn-Si}$ = 2.03 {\AA} and $h_{\rm C-C}$ = 0.21 {\AA}) of a previous LDA calculation~\cite{gla,note1}. The values of $h_{\rm Sn-Si}$ and $h_{\rm C-C}$ slightly change by less than 0.05 {\AA}, depending on the employed exchange-correlation functionals. Figures 2(a) and 2(b) show the LDA band structure and partial density of states (PDOS) projected onto the Sn 5$p$ and substrate Si 3$p$ and C 2$p$ orbitals, respectively. Interestingly, we find that Sn DB electrons form three surface bands designated as $S_1$, $S_2$, and $S_3$ [see Fig. 2(a)]. The bands projected onto the Sn 5$p_x$, 5$p_y$, and 5$p_z$ orbitals obviously indicate that $S_1$ originates from the 5$p_z$ orbital while $S_2$ and $S_3$ have mixed 5$p_x$ and 5$p_y$ characters (see Fig. 1S of the Supplemental Material~\cite{supp}). The higher energy of the $S_1$ state relative to the almost degenerate $S_2$ and $S_3$ states can be attributed to the effect of crystal-field splitting: i.e., the electrostatic repulsion between the Sn 5$p_z$ and Si 3$p_z$ orbitals is likely larger than that between the Sn 5$p_x$ (or 5$p_y$) and Si 3$p_z$ orbitals. It is noted that, for the $S_1$ state, the Sn 5$p_z$ PDOS is nearly equal in magnitude to the sum of the PDOS of Si 3$p$ and C 2$p$ orbitals, while, for the $S_2$ and $S_3$ states, the Sn 5$p_x$ + 5$p_y$ PDOS only amounts to ${\sim}$65\% of the sum of the PDOS of Si 3$p$ and C 2$p$ orbitals [see Fig. 2(b) and Fig. 2S of the Supplemental Material~\cite{supp}]. Such a large hybridization between the Sn DB state and the substrate Si and C states is well reflected by the conspicuously mixed charge character of the localized Sn-DB and the delocalized SiC(0001)-substrate electrons [see the inset of Fig. 2(a)].

\begin{figure}[ht]
\centering{ \includegraphics[width=8.0cm]{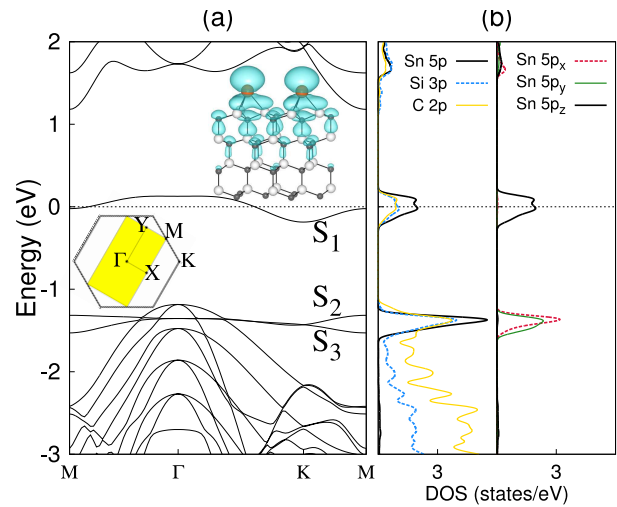} }
\caption{(Color online) (a) Surface band structure and (b) PDOS of the NM ${\sqrt{3}}{\times}{\sqrt{3}}$ structure, obtained using the LDA functional. The Brillouin zones of the NM ${\sqrt{3}}{\times}{\sqrt{3}}$ and AFM 2${\sqrt{3}}{\times}{\sqrt{3}}$ structures are drawn in the inset of (a). The charge character of the $S_1$ state at the ${\Gamma}$ point is also displayed with an isosurface of 0.002 $e$/{\AA}$^3$. The energy zero represents the Fermi level. In (b), the PDOS projected onto the Sn 5$p$, Si 3$p$, and C 2$p$ orbitals are shown, together with decomposition into the Sn 5$p_x$, 5$p_y$, and 5$p_z$ components. }
\end{figure}
\begin{figure*}[ht]
\centering{ \includegraphics[width=15cm]{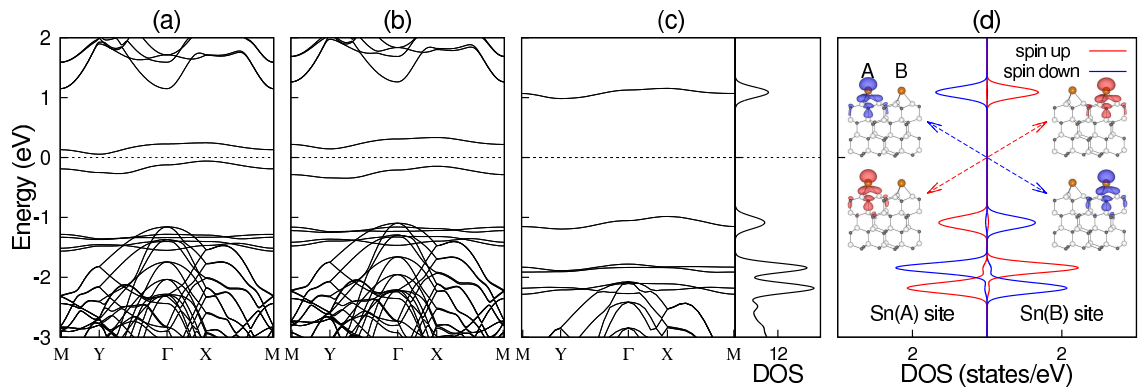} }
\caption{(Color online) Band structures of the AFM structure obtained using (a) LDA, (b) GGA, and (c) hybrid DFT. The band dispersions are plotted along the symmetry lines in the surface Brillouin zone of the unit cell [see the inset in Fig. 2(a)]. In (c), the DOS is also given. The spin-polarized local DOS projected onto the two Sn atoms at the A and B sites within the AFM structure, obtained using hybrid DFT, are given in (d). Here, the charge characters of the spin-up (spin-down) states for the highest occupied and the lowest unoccupied bands are taken at the ${\Gamma}$ point with an isosurface of 0.002 ($-$0.002) $e$/{\AA}$^3$.}
\end{figure*}

As shown in Fig. 2(a), the $S_1$ state crosses the Fermi level $E_F$, producing a half-filled band. Despite its delocalized charge character as mentioned above, the $S_1$ state has a small band width of 0.31, 0.33, and 0.55 eV, obtained using the LDA, GGA, and hybrid DFT calculations, respectively. This flat-band-like feature is likely to be attributed to a large separation of ${\sim}$5.3 {\AA} between Sn atoms within the ${\sqrt{3}}{\times}{\sqrt{3}}$ unit cell. Because of such a narrow band width of the $S_1$ state, the electronic instabilities such as a charge or spin density wave (CDW/SDW) may be expected. For the CDW instability, we find that the 3${\times}$3 structure containing three Sn atoms (i.e., U$_1$, U$_2$, and D atoms in Fig. 3S of the Supplemental Material~\cite{supp}) of different heights is more stable than the NM ${\sqrt{3}}{\times}{\sqrt{3}}$ structure by 13.4, 19.0, and 182.8 meV per ${\sqrt{3}}{\times}{\sqrt{3}}$ unit cell for LDA, GGA, and hybrid DFT, respectively (see Table I).
\begin{table}[ht]
\caption{Total energies (in meV per ${\sqrt{3}}{\times}{\sqrt{3}}$ unit cell) of the FM and AFM structures relative to the NM structure, calculated using LDA, GGA and hybrid DFT. }
\begin{ruledtabular}
\begin{tabular}{lrrr}
   & CDW & FM & AFM     \\ \hline
LDA & $-$13.4  & $-$16.1 & $-$28.7   \\
GGA & $-$19.0  &  $-$61.8 & $-$69.3   \\
hybrid DFT & $-$182.8   & $-$410.8 & $-$446.5  \\
\end{tabular}
\end{ruledtabular}
\end{table}
Since such a buckled NM 3${\times}$3 structure accompanies a charge transfer from the D to the U$_1$ (or U$_2$) atoms, it is most likely to reduce Coulomb repulsions between Sn DB electrons compared to the NM ${\sqrt{3}}{\times}{\sqrt{3}}$ structure. We note that the calculated band structure of the NM 3${\times}$3 structure exhibits the presence of occupied surface states at $E_F$ (see Fig. 4S of the Supplemental Material~\cite{supp}), indicating a metallic feature. To find the possibility of SDW, we perform the spin-polarized LDA, GGA, and hybrid DFT calculations for the ferromagnetic (FM) ${\sqrt{3}}{\times}{\sqrt{3}}$ and AFM 2${\sqrt{3}}{\times}{\sqrt{3}}$ structures, which were considered in the previous LDA calculation~\cite{gla}. We find that all of the employed exchange-correlation functionals favor the FM and AFM structures over the NM ${\sqrt{3}}{\times}{\sqrt{3}}$ and 3${\times}$3 structures (see Table I). Here, the AFM structure is more stable than the FM structure, consistent with the previous LDA calculation~\cite{gla}. It is noted that the stabilities of the two magnetic structures relative to the NM ${\sqrt{3}}{\times}{\sqrt{3}}$ structure increase in the order of LDA $<$ GGA $<$ hybrid DFT calculations (see Table I). In the optimized AFM structure, two Sn atoms within the 2${\sqrt{3}}{\times}{\sqrt{3}}$ unit cell are at the same height, indicating a ${\sqrt{3}}{\times}{\sqrt{3}}$ structural symmetry as observed by low-energy electron diffraction and scanning tunneling microscopy~\cite{gla}.

Figures 3(a), 3(b), and 3(c) show the LDA, GGA, and hybrid-DFT band structures of the AFM structure, which give the band gap $E_g$ of 0.12, 0.30, and 1.97 eV, respectively. The band gap obtained using hybrid DFT is found to be closer to that (${\sim}$2 eV) measured by photoemission spectroscopy~\cite{gla}. As shown in Fig. 3(c), the DOS obtained using hybrid DFT exhibits the three peaks located at $-$1.09, $-$1.83, and $-$2.17 eV below $E_F$, which are associated with the $S_1$, $S_2$, and $S_3$ states, respectively. On the other hand, photoemission spectra~\cite{gla} showed the presence of two peaks at $-$1.0 and $-$2.4 eV, which were interpreted to originate from the Sn DB state and the SiC bulk states, respectively. Based on the present DOS results, we however interpret the upper and lower photoemission peaks in terms of the $S_1$ and $S_2$ (or $S_3$) surface states, respectively.

To understand the underlying mechanism for the gap opening of the AFM spin ordering, we plot in Fig. 3(d) the spin-polarized local DOS projected onto the two Sn atoms at A and B sites, together with their spin characters. It is seen that the occupied (unoccupied) spin-up and spin-down $S_1$ states are localized at the A(B) and B(A) sites, respectively. Here, the hybridization between the occupied spin-up (spin-down) state at the A(B) site and the unoccupied spin-up (spin-down) state at the B(A) site gives rise to a gap opening~\cite{sato}. Such superexchange interaction~\cite{super1,super2} between the occupied and unoccupied electronic states can be facilitated due to a large hybridization of the Sn 5$p_z$ orbitals with the Si 3$p_z$ and C 2$p_z$ orbitals [see Fig. 2(b) and Fig. 2S of the Supplemental Material~\cite{supp}]. This superexchange interaction is well represented by a large spin delocalization [see Fig. 3(d)] with the spin moments of ${\pm}$0.33, ${\pm}$0.12, and ${\pm}$0.10 ${\mu}_B$ for Sn, Si (outermost-layer) and C (outermost-layer) atoms, respectively (see the hybrid DFT results in Table IS of the Supplemental Material~\cite{supp}). We note that the calculated spin moments of Sn, Si, and C atoms increase in the order of LDA $<$ GGA $<$ hybrid DFT calculations (see Table IS), corresponding to that of the stabilization energy of the AFM structure (see Table I). On the basis of our DFT calculations, we can say that the magnetically driven insulating state of Sn/SiC(0001) with a large spin delocalization can be characterized as a Slater-type insulator.

The existence of the long-range AFM order due to the sizable hybridization between the Sn DB state and the substrate states raises questions about the reliability of the previous LDA + DMFT study~\cite{gla} in which a single-band Hubbard model including only the on-site Coulomb interaction was employed. Here, the single band representing the DB state dominantly localized at Sn atoms invokes strong on-site Coulomb repulsion with suppressed electron hoping, driving the gap formation. Despite the fact that such a model Hamiltonian does not incorporate long-range interactions due to the largely hybridized $S_1$ state, we solve it within the LDA + DMFT scheme~\cite{DMFT,dmftref,qmc}. Figure 4(a) shows the calculated DOS for the AFM and paramagnetic phases obtained at $T$ = 100 and 300 K, respectively. The observed insulating gap of ${\sim}$2 eV is found to be well reproduced with $U$ = 1.8 eV, similar to the previous~\cite{gla} LDA + DMFT calculation. As shown in Fig. 4(b), the paramagnetic phase is transformed into the AFM phase below $T_N$ ${\approx}$ 100 K. Note that such a phase transition little changes the insulating gap [see Fig. 4(a)]. Therefore, the LDA + DMFT results indicate that the gap formation is not driven by the AFM order but attributed to the on-site interaction, representing a Mott-type insulator. Accordingly, the spin magnetic moment obtained using LDA + DMFT is 1 ${\mu}_B$ for Sn atom [see Fig. 4(b)]. Such a localized magnetic moment inherent in the Mott phase drastically contrasts with the large spin delocalization over Sn atoms and Si substrate atoms obtained using the hybrid DFT calculation [see Fig. 3(d) and Table IS]. Future experiments are anticipated to resolve such different features of spin magnetic moment between the Mott-type and Slater-type insulators by measuring the surface magnetic moments at Sn/SiC(0001).

\begin{figure}[ht]
\centering{ \includegraphics[width=8.0cm]{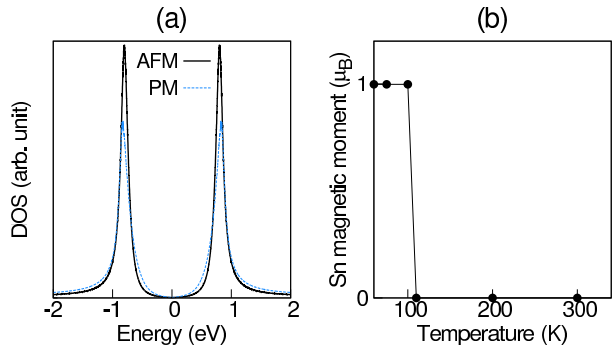} }
\caption{(Color online) (a) Density of states of the AFM and paramagnetic (PM) phases at 100 and 300 K, respectively, obtained using the LDA + DMFT calculation with $U$ = 1.8 eV. The calculated spin magnetic moment of Sn atom is plotted as a function of temperature in (b).}
\end{figure}

It is noteworthy that the charge character of the $S_1$ state exhibits a large delocalization up to the third deeper Si and C substrate layers, which in turn gives some lateral overlap between neighboring Sn atoms [see the inset of Fig. 2(a)]. Such an extension of the half-filled surface state calls for the importance of long-range interactions which were not considered in the previous~\cite{gla} and present LDA + DMFT calculations. Indeed, a recent fully self-consistent GW + DMFT study~\cite{hans} for the analogous X/Si(111) systems (with X = C, Si, Sn, and Pb) reported that taking into account long-range Coulomb interactions is mandatory because of their comparable magnitude with that of the on-site Coulomb interaction. It was shown that the inclusion of long-range interactions within the extended Hubbard model changes the ground-state character of the X/Si(111) systems~\cite{hans}: i.e., without long-range interactions, all the X/Si(111) systems are in the Mott phase, but, as long-range interactions are added, Sn/Si(111) and Pb/Si(111) become closer to a metallic phase. Compared to the Sn/Si(111) system, Sn/SiC(0001) has the ${\sim}$20\% smaller nearest-neighbor distance of Sn atoms as well as the relatively lower dielectric screening of the SiC substrate, thereby leading to an increase in the intersite interactions. It is thus expected that the nonlocal interaction effects in Sn/SiC(0001) might significantly influence the stability of the Mott phase obtained by using only the on-site interaction. For more accurate simulation of the present system, the extended Hubbard model including long-range interaction terms will be demanded in future theoretical work. There still remains an interesting challenge of how to equally consider all of the on-site interaction, long-range interactions, and magnetic response in the Sn/SiC(0001) system.

To conclude, we have presented two different pictures for the insulating nature of the Sn overlayer on a wide-gap SiC(0001) substrate using the LDA, GGA, and hybrid DFT calculations and the LDA + DMFT calculation. The DFT calculations drew the Slater-type picture with a long-range AFM order, while the LDA + DMFT calculation supported the Mott-type picture driven by strong on-site Coulomb repulsion. Unexpectedly, the Sn DB state was found to largely hybridize with the substrate Si and C states, thereby facilitating the stabilization of the AFM spin ordering via superexchange interactions. This intriguing electronic structure of the present system raises an important issue of how long-range interactions beyond the on-site interaction should be taken into account to diminish the Mott phase. Our findings will not only caution against the realization of the Mott-insulating phase in metal overlayers on semiconductor substrates, but also stimulate further experimental studies for the exploration of the magnetic phases of Sn/SiC(0001).

\noindent {\bf Acknowledgement.}
This work was supported by National Research Foundation of Korea (NRF) grant funded by the Korea Government (MSIP) (2015R1A2A2A01003248). The calculations were performed by KISTI supercomputing center through the strategic support program (KSC-2013-C3-043) for the supercomputing application research.

\noindent $^{*}$ Corresponding author: chojh@hanyang.ac.kr

\widetext
\clearpage

\makeatletter
\renewcommand{\fnum@figure}{\figurename ~\thefigure{S}}
\renewcommand{\fnum@table}{\tablename ~\thetable{S}}
\makeatother

\vspace{2.4cm}
{\bf \huge Supplemental Material}
\vspace{0.2cm}

\vspace{1cm}
{\bf \large 1. Sn $p$-orbitals projected band structure of the NM structure}
\vspace{0.2cm}

To figure out the electronic characters of the $S_1$, $S_2$, and $S_3$ states, we plot in Fig. 1S the bands projected onto the Sn 5$p_x$, 5$p_y$, and 5$p_z$ orbitals. It is shown that $S_1$ originates from the 5$p_z$ orbital while $S_2$ and $S_3$ have mixed 5$p_x$ and 5$p_y$ characters.

\begin{figure}[h]
\centering{ \includegraphics[width=10.0cm]{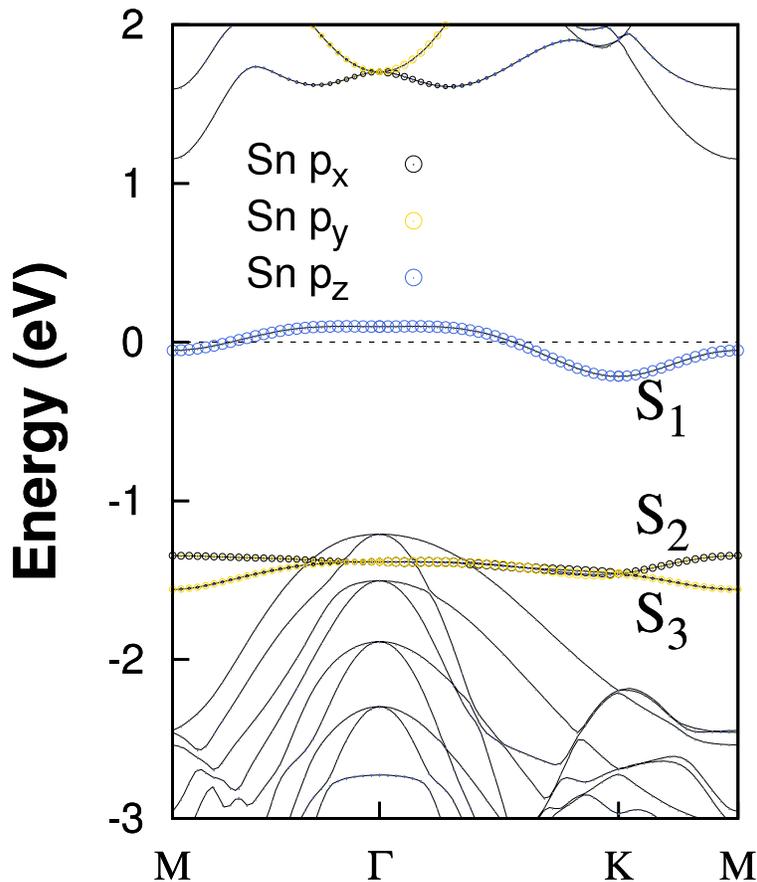} }
\caption{Calculated NM bands projected onto the Sn 5$p_x$, 5$p_y$, and 5$p_z$ orbitals, obtained using the LDA calculation. }
\end{figure}

\newpage
{\bf \large 2. Partial density of states of the NM structure}
\vspace{0.2cm}

Figure 2S shows the partial density of states (PDOS) of the NM structure, projected onto the Si 3$p_x$, 3$p_y$, and 3$p_z$ components as well as the C 2$p_x$, 2$p_y$, and 2$p_z$ ones.

\begin{figure}[h]
\centering{ \includegraphics[width=14.0cm]{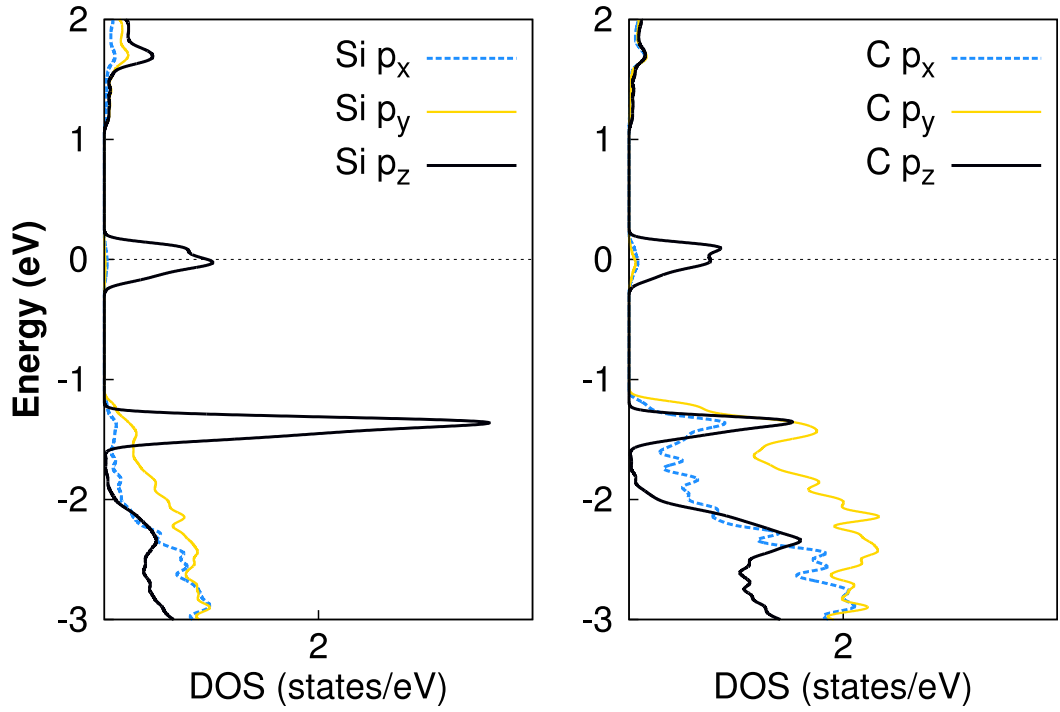} }
\caption{(Color online) Calculated PDOS of the NM ${\sqrt{3}}{\times}{\sqrt{3}}$ structure, obtained using the LDA calculation. The PDOS projected onto the Si 3$p_x$, 3$p_y$, and 3$p_z$ (C 2$p_x$, 2$p_y$, and 2$p_z$) orbitals is shown in the left (right) panel.}
\end{figure}

\vspace{2cm}

\newpage
{\bf \large 3. Geometry of the NM 3${\times}$3 structure}
\vspace{0.2cm}
\begin{figure}[h]
\centering{ \includegraphics[width=14.0cm]{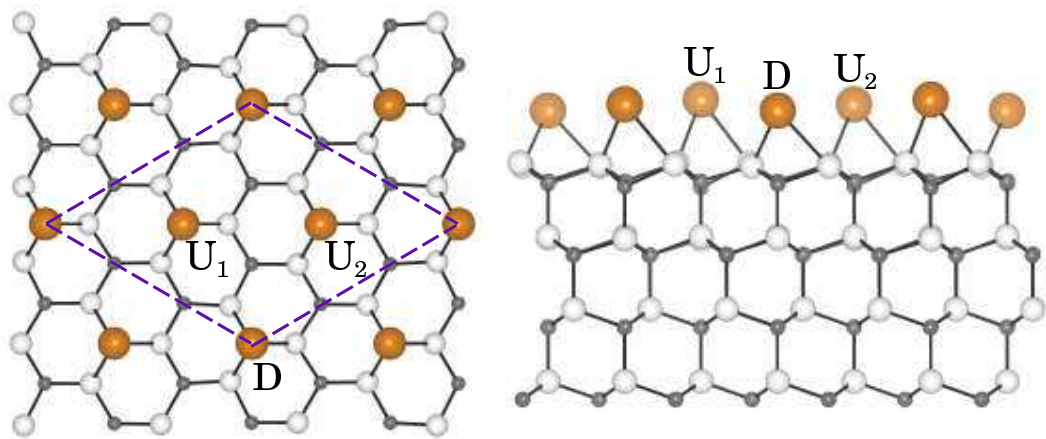} }
\caption{(Color online) Top (left) and side (right) views of the NM 3${\times}$3 structure. The dashed line indicates the ${3}{\times}{3}$ unit cell, where three Sn atoms of different heights are designated as U$_1$, U$_2$, and D. Here the position of U$_1$ (U$_2$) is higher than that of D by 0.38 (0.22) {\AA}.  }
\end{figure}

\newpage
{\bf \large 4. Band structure of the NM 3${\times}$3 structure}
\vspace{0.2cm}

Figure 4S shows the calculated band structure of the NM 3${\times}$3 structure. It is seen that there is a half-filled band crossing the Fermi level, indicating a metallic feature.

\begin{figure}[h]
\centering{ \includegraphics[width=14.0cm]{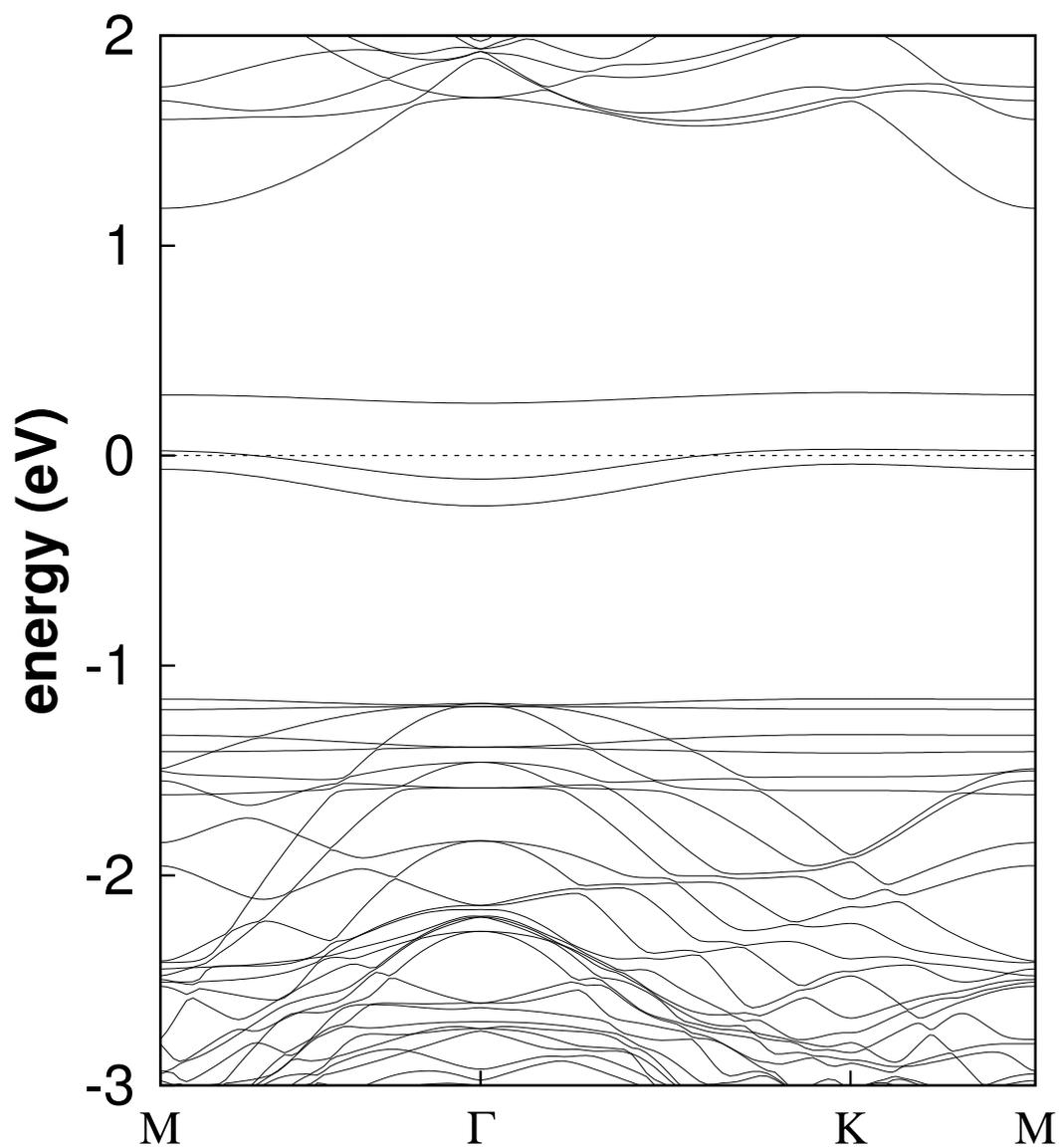} }
\caption{ Calculated band structure of the NM 3${\times}$3 structure obtained using the LDA calculation. The energy zero represents the Fermi level.}
\end{figure}

\newpage
{\bf \large 5. Spin moments of Sn, Si, and C atoms in the AFM structure}
\vspace{0.2cm}
\begin{table}[ht]
\caption{Calculated spin magnetic moments (in $\mu_B$) of the AFM structure within the PAW spheres centered at Sn, Si, and C atoms, obtained using the HSE calculation. The PAW sphere radii were chosen as 1.566, 1.312, and 0.863 {\AA} for Sn, Si, and C atoms, respectively. The numbering of each atom is seen in the figure below. The results obtained by using the LDA and GGA calculations are also given in parentheses as($m_{LDA}$, $m_{GGA}$). }
\begin{ruledtabular}
\begin{tabular}{lcccccc}
Sn atoms  &             &     Sn$_{1}$&             &                   &           Sn$_{2}$&                   \\ \hline
          &             &        0.334&             &                   &           $-$0.334&                   \\
          &             &(0.284,0.291)&             &                   &($-$0.284,$-$0.291)&                   \\
1st layer &     Si$_{1}$&     Si$_{2}$&     Si$_{3}$&    Si$_{1^\prime}$&    Si$_{2^\prime}$&    Si$_{3^\prime}$\\s
          &        0.039&        0.039&        0.039&           $-$0.039&           $-$0.039&           $-$0.039\\
          &(0.029,0.031)&(0.029,0.031)&(0.029,0.031)&($-$0.029,$-$0.031)&($-$0.029,$-$0.031)&($-$0.029,$-$0.031)\\
2ed layer &      C$_{1}$&      C$_{2}$&      C$_{3}$&     C$_{1^\prime}$&     C$_{2^\prime}$&     C$_{3^\prime}$\\
          &        0.004&        0.091&        0.004&           $-$0.004&           $-$0.091&           $-$0.004\\
          &(0.004,0.004)&(0.081,0.086)&(0.004,0.004)&($-$0.004,$-$0.004)&($-$0.081,$-$0.086)&($-$0.004,$-$0.004)\\
3rd layer &     Si$_{1}$&     Si$_{2}$&     Si$_{3}$&    Si$_{1^\prime}$&    Si$_{2^\prime}$&    Si$_{3^\prime}$\\
          &        0.001&        0.037&        0.001&           $-$0.001&           $-$0.037&           $-$0.001\\
          &(0.001,0.001)&(0.033,0.036)&(0.001,0.001)&($-$0.001,$-$0.001)&($-$0.033,$-$0.036)&($-$0.001,$-$0.001)\\
4th layer &      C$_{1}$&      C$_{2}$&      C$_{3}$&     C$_{1^\prime}$&     C$_{2^\prime}$&     C$_{3^\prime}$\\
          &        0.003&        0.003&        0.003&           $-$0.003&           $-$0.003&           $-$0.003\\
          &(0.002,0.003)&(0.002,0.003)&(0.002,0.003)&($-$0.002,$-$0.003)&($-$0.002,$-$0.003)&($-$0.002,$-$0.003)\\
\end{tabular}
\end{ruledtabular}
\end{table}
\begin{figure}[h]
\centering{ \includegraphics[width=10.0cm]{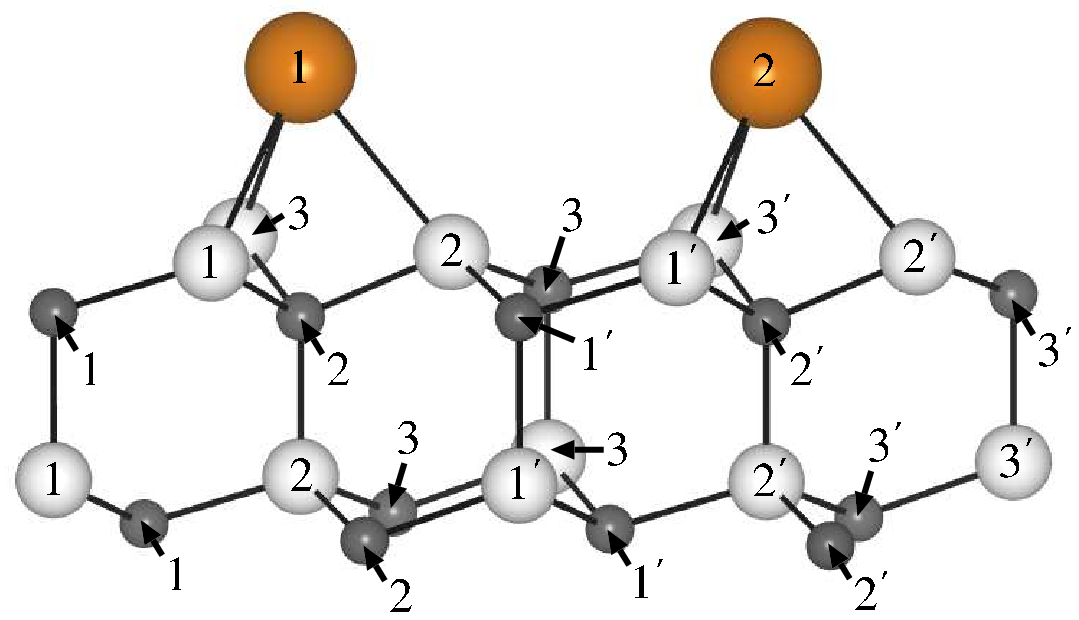} }
\end{figure}


\begin{thebibliography}{99}

\bibitem{ima} M. Imada, A. Fujimori, and Y. Tokura, Rev. Mod. Phys. {\bf 70}. 1039 (1998)
\bibitem{meng} Z. Y. Meng, T. C. Lang, S. Wessel, F. F. Assaad, and A. Muramatsu, Nature {\bf 464}, 847 (2010).
\bibitem{nor} J. E. Northrup and J. Neugebauer, Phys. Rev. B. {\bf 57}, 4230(R) (1998)
\bibitem{ani} V. I. Anisimov, A. E. Bedin, M. A. Korotin, G. Santoro, S. Scandolo, and E. Tosatti, Phys. Rev. B {\bf 61}, 1752 (2000)
\bibitem{cor} R. Cort\'es, A. Tejeda, J. Lobo, C. Didiot, B. Kierren, D.Malterre, E. G. Michel, and A. Mascaraque, Phys. Rev. Lett. {\bf 96}, 126103 (2006).
\bibitem{mod} S. Modesti, L. Petaccia, G. Ceballos, I. Vobornik, G. Panaccione, G. Rossi, L. Ottaviano, R. Larciprete, S. Lizzit, and A. Goldoni, Phys. Rev. Lett. {\bf 98}, 126401 (2007).
\bibitem{pro} G. Profeta and E. Tosatti, Phys. Rev. Lett. {\bf 98}, 086401 (2007).
\bibitem{li} G. Li, M. Laubach, A. Fleszar, and W. Hanke, Phys. Rev. B {\bf 83}, 041104(R) (2011).
\bibitem{li2} G. Li, P. H\"{o}pfner, J. Sch\"{a}fer, C. Blumenstein, S. Meyer, A. Bostwick, E. Rotenberg, R. Claessen, and W. Hanke, Nat. Commun. {\bf 4}, 1620 (2013).
\bibitem{mor} H. Morikawa, I. Matsuda, and S. Hasegawa, Phys. Rev. B {\bf 65}, 201308(R) (2002).
\bibitem{car} J. M. Carpinelli, H. H. Weitering, M. Bartkowiak, R. Stumpf, and E. W. Plummer, Phys. Rev. Lett. {\bf 79}, 2859 (1997).
\bibitem{bal} G. Ballabio, S. Scandolo and E. Tosatti, Phys. Rev. B {\bf 61}, 13345(R) (2000).
\bibitem{lee1} J.-H. Lee, H.-J. Kim, and J.-H. Cho, Phys. Rev. Lett. {\bf 111}, 106403 (2013).
\bibitem{lee2} J.-H. Lee, X.-Y. Ren, Y. Jia, and J.-H. Cho, Phys. Rev. B {\bf 90}, 125439 (2014).
\bibitem{gla} S. Glass, G. Li, F. Adler, J. Aulbach, A. Fleszar, R. Thomale, W. Hanke, R. Claessen, and J. Sch\"{a}fer, Phys. Rev. Lett. {\bf 114}, 247602 (2015).
\bibitem{method} We have performed the LDA, GGA, and hybrid DFT calculations using the Vienna $ab$ $initio$ simulation package (VASP) with the projector augmented wave method [Kresse and Hafner, Phys. Rev. B {\bf 48,} 13115 (1993); Kresse and Furthm\"{u}ller, Comput. Mater. Sci. {\bf 6,} 15 (1996)]. For the exchange-correlation energy, we employed the LDA functional of Ceperley-Alder (CA) [Ceperley and Alder, Phys. Rev. Lett. {\bf 45,} 566 (1980)], the GGA functional of Perdew-Burke-Ernzerhof (PBE) [Perdew, Burke, and Ernzerhof, Phys. Rev. Lett. {\bf 77,} 3865 (1996);{\bf 78,} 1396(E) (1997)], and the hybrid functional of Heyd-Scuseria-Ernzerhof (HSE) [Heyd, Scuseria, and Erzerhof, J. Chem. Phys. {\bf 118}, 8207 (2003); Krukau $et$ $al.$, J. Chem. Phys. {\bf 125}, 224106 (2006)]. Since the HSE functional with a mixing factor of ${\alpha}$ = 0.5 controlling the amount of exact Fock exchange energy predicts well the observed insulating gap of ${\sim}$2 eV, we used this optimal ${\alpha}$ value for the hybrid DFT calculation. The SiC(0001) substrate [with the optimized lattice constant $a_0$ = 3.063 (3.097) {\AA} for the LDA (GGA) calculation] was modeled by a periodic slab geometry consisting of the eight-layer slab with ${\sim}$20 {\AA} of vacuum in between the slabs. For the hybrid DFT calculation, we used the lattice constant optimized by the GGA calculation. Each C atom in the bottom layer of the slab was passivated by one H atom. The ${\bf k}$-space integrations for the nonmagnetic (or FM) and AFM structures were done with the ${\Gamma}$-centered 18${\times}$18 and 9${\times}$18 uniform meshes in the surface Brillouin zones of the ${\sqrt{3}}{\times}{\sqrt{3}}$ and 2${\sqrt{3}}{\times}{\sqrt{3}}$ unit cells, respectively. All atoms except the bottom Si and C layers were allowed to relax along the calculated forces until all the residual force components were less than 0.01 eV/{\AA}.
\bibitem{note1} The previous LDA calculation of Glass $et$ $al$.~\cite{gla} employed a periodic slab geometry consisting of the six-layer slab, while the present DFT calculation used the eight-layer slab.
\bibitem{supp} See Supplemental Material at http://link.aps.org/supplemental/xxxx for the Sn $p$-orbitals projected band structure and PDOS of the NM structure, the geometry and band structure of the NM 3${\times}$3 structure, and the spin magnetic moments of Sn, Si, and C atoms in the AFM structure.
\bibitem{sato} K. Sato, L. Bergqvist, J. Kudrnovsky, P. H. Dederichs, O. Eriksson, I. Turek, B. Sanyal,G. Bouzerar,H. Katayama-Yoshida, V. A. Dinh, T. Fukushima, H. Kizaki, and R. Zeller, Rev. Mod. Phys. {\bf 82}, 1633 (2010).
\bibitem{super1} J. B. Goodenough, Phys. Rev. {\bf 100}, 564 (2010).
\bibitem{super2} J. Kanamori, J. Phys. Chem. Solids {\bf 10}, 87 (1959).
\bibitem{DMFT} The low-energy effective single-band Hubbard model with on-site Coulomb repulsion $U$ is given by
\begin{equation}
H = \sum_{\rm{k},\sigma} \epsilon_{\rm{k},\sigma}^{\text{LDA}} + \sum_j U n_{j,\uparrow} n_{j,\downarrow},
\end{equation}
where $\epsilon_{\rm{k},\sigma}^{\text{LDA}}$ are the LDA eigenvalue of the $S_1$ state and $n_{j,{\sigma}}$ is the
number operator counting electrons at the Sn-atom site $j$. The DMFT self-consistency equation reads
\begin{eqnarray}
G_{j,\sigma} (i\omega_n) =
\int d\epsilon_k \frac{1}{\begin{pmatrix} i\omega_n + \mu - \Sigma_{A,\sigma} (i\omega_n) & \epsilon_k \\
\epsilon_k & i\omega_n + \mu - \Sigma_{B,\sigma} (i\omega_n) \end{pmatrix}}, \nonumber
\end{eqnarray}
where $\mu$ is the chemical potential, $\omega_n$ is the Matsubara frequency, and $j$ = $A,B$. Here, we employed the continuous-time
quantum Monte Carlo approach as an impurity solver.
\bibitem{dmftref} A. Georges, G. Kotliar, W. Krauth, and M.J. Rozenberg, Rev. Mod. Phys. {\bf 68}, 13 (1996).
\bibitem{qmc} E. Gull, A. J. Millis, A. I. Lichtenstein, A. N. Rubtsov, M. Troyer, and P. Werner, Rev. Mod. Phys. {\bf 83}, 349 (2011).
\bibitem{hans} P. Hansmann, T. Ayral, L. Vaugier, P. Werner, and S. Biermann, Phys. Rev. Lett. {\bf 110}, 166401 (2013).
\end{thebibliography}
\end{document}